\UseRawInputEncoding

\documentclass[floatfix,aps,prb,twocolumn,superscriptaddress]{revtex4-2}

\usepackage[table]{xcolor}
\usepackage{graphicx}
\usepackage{dcolumn}
\usepackage{bm}
\usepackage{hyperref}
\usepackage{float}

\makeatother

\begin{document}

\preprint{APS/123-QED}

\title{VASPilot: MCP-Facilitated Multi-Agent Intelligence for Autonomous VASP Simulations}

\author{Jiaxuan Liu}
\author{Tiannian Zhu}
\author{Caiyuan Ye}
\affiliation{%
Beijing National Laboratory for Condensed Matter Physics and Institute of Physics,
Chinese Academy of Sciences, Beijing 100190, China
}
\affiliation{%
University of Chinese Academy of Sciences, Beijing 100049, China
}

\author{Zhong Fang}
\author{Hongming Weng}
\email{hmweng@iphy.ac.cn} 
\affiliation{%
Beijing National Laboratory for Condensed Matter Physics and Institute of Physics,
Chinese Academy of Sciences, Beijing 100190, China
}
\affiliation{%
University of Chinese Academy of Sciences, Beijing 100049, China
}

\affiliation{%
Songshan Lake Materials Laboratory, Dongguan, Guangdong 523808, China
}
\author{Quansheng Wu}
\email{quansheng.wu@iphy.ac.cn} 
\affiliation{%
Beijing National Laboratory for Condensed Matter Physics and Institute of Physics,
Chinese Academy of Sciences, Beijing 100190, China
}
\affiliation{%
University of Chinese Academy of Sciences, Beijing 100049, China
}

\date{\today}

\begin{abstract}
Density-functional-theory (DFT) simulations with the Vienna Ab initio Simulation Package (VASP) are indispensable in computational materials science but often require extensive manual setup, monitoring, and postprocessing. Here, we introduce VASPilot, an open-source platform that fully automates VASP workflows via a multi-agent architecture built on the CrewAI framework and a standardized Model Context Protocol (MCP). VASPilot's agent suite handles every stage of a VASP study-from retrieving crystal structures and generating input files to submitting Slurm jobs, parsing error messages, and dynamically adjusting parameters for seamless restarts. A lightweight Flask-based web interface provides intuitive task submission, real-time progress tracking, and drill-down access to execution logs, structure visualizations, and plots. We validate VASPilot on both routine and advanced benchmarks: automated band-structure and density-of-states calculations (including on-the-fly symmetry corrections), plane-wave cutoff convergence tests, lattice-constant optimizations with various van der Waals corrections, and cross-material band-gap comparisons for transition-metal dichalcogenides. In all cases, VASPilot completed the missions reliably and without manual intervention. Moreover, its modular design allows easy extension to other DFT codes simply by deploying the appropriate MCP server. By offloading technical overhead, VASPilot enables researchers to focus on scientific discovery and accelerates high-throughput computational materials research.

\end{abstract}

\maketitle

\section{\label{sec:intro}Introduction}

With the advancement of computer hardware and algorithm, density functional theory (DFT) calculations have become one of the most important research tools in materials science, chemistry, and condensed matter physics. Starting from quantum mechanics, DFT can predict the mechanical, electromagnetic, and optical properties of materials, playing a crucial role in areas such as batteries\cite{doi:10.1021/cm3027219, wang_zero-strain_2013, wang_oxysulfide_2017}, topological phenomena\cite{jia_moire_2024, wang_fractional_2024, devakul_magic_2021} and transport\cite{zhang_magnetoresistance_2019, liu_first-principles_2024, pi_first-principles_2024}.

Researchers have developed numerous DFT software packages based on different basis sets and treatments of core electrons. For example, Vienna Ab initio Simulation Package (VASP)\cite{kresse_efficient_1996}, Quantum Espresso\cite{giannozzi_quantum_2009, giannozzi_advanced_2017}, and CASTEP\cite{clark_first_2005} employ plane-wave basis sets, while OpenMX\cite{ozaki_variationally_2003, weng_revisiting_2009} and Siesta\cite{soler_siesta_2002} are based on atomic orbital basis sets. Additionally, there are codes like WIEN2k\cite{blaha_wien2k_2020} and Elk\cite{noauthor_elk_nodate} that utilize all-electron wavefunctions in calculations. Although these software are widely used in various research fields, researchers still need to invest considerable effort in the complicated workflow of these software.

VASPKIT\cite{WANG2021108033}, ASE\cite{hjorth_larsen_atomic_2017}, and Pymatgen\cite{ong_python_2013} have alleviated this pain point to a certain extent. VASPKIT is a command-line-based software that can quickly generate corresponding input files for calculations based on crystal structures upon user's command. Moreover, it facilitates rapid plotting using predefined templates after the computations are completed. ASE is a Python-based molecular dynamics simulation software that integrates interfaces with various DFT packages such as VASP, Quantum Espresso, CASTEP, and Siesta. By writing Python scripts, researchers can efficiently conduct Density Functional Theory (DFT) calculations. Furthermore, ASE supports interfaces with machine learning interatomic potentials software like DeepMD\cite{wang_deepmd-kit_2018}, 
NequIP\cite{batzner_e3-equivariant_2022}, and CHGNET\cite{deng_chgnet_2023}. This capability enables rapid and accurate predictions of mechanical properties, bridging the gap between traditional DFT calculations and large-scale atomistic simulations. Pymatgen is a Python-based high-throughput computational materials science library with extensive interfaces for VASP. Through tailored Python scripting, researchers can efficiently prepare input files and seamlessly import calculation results into Python for data analysis and visualization. 

These robust interfaces have made high-throughput Density Functional Theory (DFT) calculations a practical reality. For repetitive tasks, these tools, when used in conjunction with custom scripts, can drastically cut down on redundant work. However, in practical research scenarios, scientists often encounter far more intricate challenges. They may need to analyze certain properties across different materials or compare the behavior of a single material under varying conditions and parameters. Such diverse and complex computational tasks demand not only the preparation of various input files tailored to each specific case but also the repetitive submission of jobs and waiting for their completion. This process can be time-consuming and labor-intensive, highlighting the necessity for further advancements in automation and integration within computational workflows. Automatic execution and management of these tasks will allow researchers to focus more on scientific discovery rather than on the fussy details of computation.

In recent years, the rapid advancement of large language models (LLMs)\cite{yang_qwen3_2025, deepseek-ai_deepseek-v3_2025, deepseek-ai_deepseek-r1_2025, team_glm-41v-thinking_2025, Meta2024Llama4, team_gemma_2025} and agent-based technologies has opened new possibilities for building automatic DFT calculation schema. Large language models are deep neural networks pretrained on vast amounts of general textual corpora and further fine-tuned on dialogue data, enabling them to emulate human-like language understanding and reasoning to a significant extent. LLMs have been applied to research in chemistry and material science. Some use LLMs as predictive model to predict the synthesizability of crystal structures\cite{doi:10.1021/jacs.4c05840, song_is_2024} and other properties. Others utilize LLM to extract data from articles\cite{doi:10.1021/jacs.3c05819, lee_text-mined_2025}. There is also research to fine-tune LLM on data of domain-specific data\cite{wang_perovskite-r1_2025}. 

Based on conversational capabilities, LLM developers have incorporated structured input-output formats into the fine-tuning process, equipping these models with the ability to invoke external programs. This ability is known as function calling or tool use. As LLMs continue to improve in performance, agent systems augmenting LLMs by tools and context have attracted wide attention. These systems not only provide more accurate responses but also gain the ability to interact with computational tools and environments. 

Due to the accumulation of lengthy context from multiple tool invocations, single-agent systems often struggle to maintain optimal performance when executing complex, multistep tasks. To address this limitation, multi-agent frameworks such as MetaGPT\cite{hong_metagpt_2024}, LangGraph\cite{langchain_ai_langgraph_2023}, and CrewAI\cite{crewai_crewai_2024} have emerged as preferred solutions for managing complex workflows. These frameworks decompose intricate tasks into a series of simpler subtasks, which are then assigned to specialized agents equipped with distinct tools, domain knowledge, and independent contextual states. By enabling coordinated execution, such frameworks achieve improved robustness and scalability, empowering LLMs to effectively handle complex scientific computing tasks. In this way, multi agent systems have potentials to become an AI coworker of human researchers.

There have been some successful attempts to leverage LLM-based multi-agent frameworks (MAF) to assist materials research. For instance, AtomAgents\cite{ghafarollahi_atomagents_2024} aims to automate alloy design by integrating a knowledge base with LAMMPS\cite{thompson_lammps_2022}-based molecular dynamics simulation tools. TopoMAS\cite{zhang_topomas_2025} accelerates the discovery and investigation of topological materials by combining literature mining, generative models, and VASP calculations. There are attempts such as MatPilot\cite{ni_matpilot_nodate} and ChemAgent\cite{song_multiagent-driven_2025} that leverage robotic arms to oversee and conduct experimental procedures. Moreover, El Agente Q\cite{zou_agente_2025} and DREAMS\cite{wang_dreams_2025}, respectively, explore the MAF on xTB\cite{bannwarth_extended_2021} and Quantum Espresso to accelerate DFT computations. 

Despite these initial demonstrations for multi-agent frameworks (MAFs) to contribute to chemistry and materials science research, several challenges remain to be addressed for their effective use in facilitating practical computations. Firstly, beyond routine tasks like band structure and DOS calculations, MAFs should also handle complex tasks without task-specific examples to maximize its adaptability. Secondly, although MAFs can expedite research processes, the role of human scientists remains indispensable. Consequently, the user interface is crucial. Scientists need real-time, intuitive understanding of the status of MAF to ensure the validity and correctness of results. Lastly, while VASP software boasts a substantial user base within the field of condensed matter physics, there has been a notable lack of focus on developing MAFs specifically tailored for VASP, indicating an area ripe for further exploration and development. 

To bridge these gaps, in this work, we introduce an open source software, VASPilot. VASPilot leverages the CrewAI framework for multi-agent cooperation, facilitating automated DFT calculation and visualization for more complicated misssions. This tool aims to further simplify the workflow of VASP even for zero-shot mission and practically reduce the researchers' effort executing VASP calculation.

In the following, we first describe the VASPilot architecture, covering the CrewAI-based multi-agent collaboration core, the Model Context Protocol (MCP) tool server, and the web-based user interface. We then demonstrate VASPilot’s robustness and versatility through a series of benchmark studies. Finally, we present a concise conclusion and outlook.

\begin{figure}[!htbp]
    \centering
    \includegraphics[width=1\linewidth]{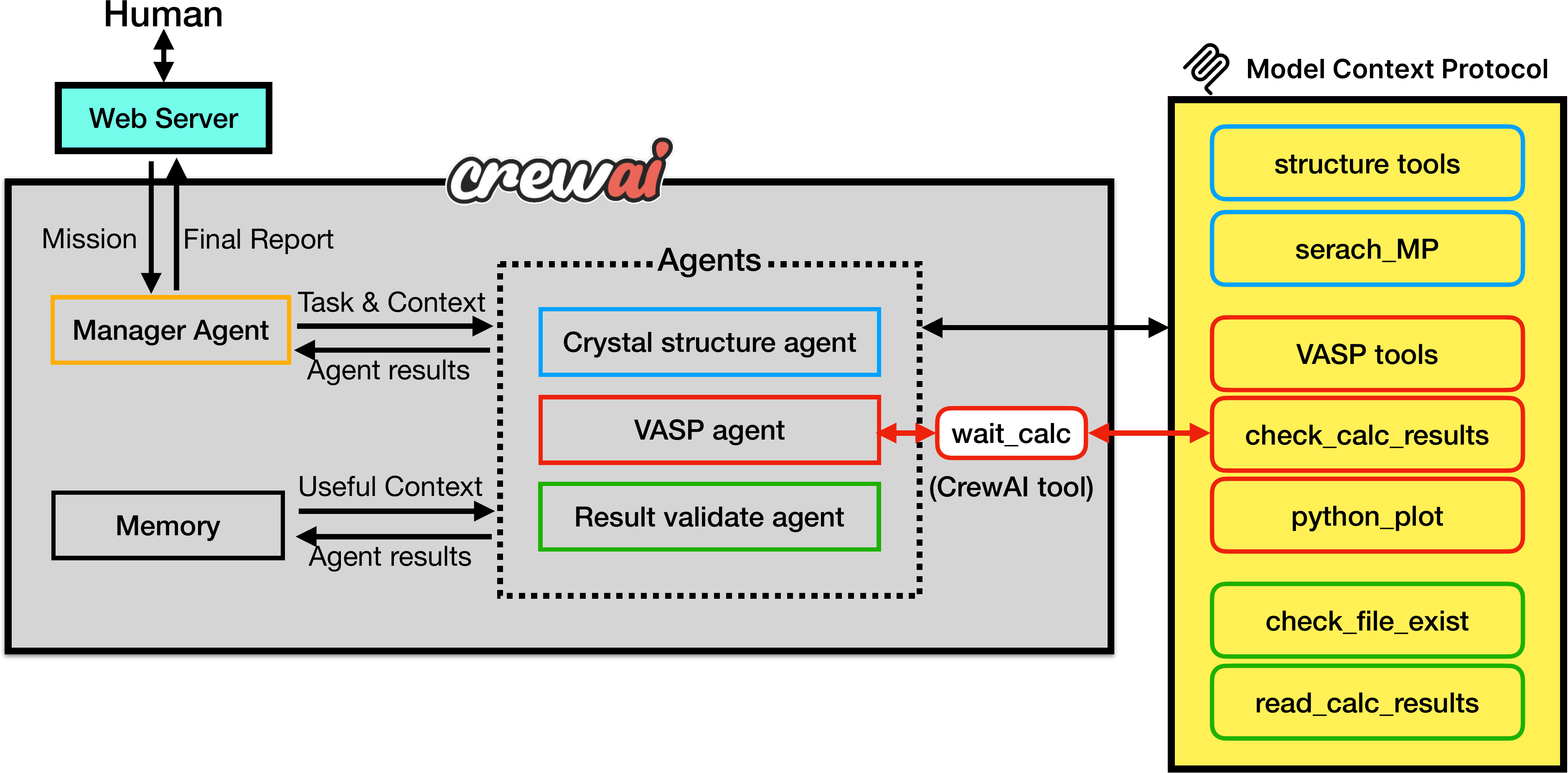}
    \caption{Overall stucture of VASPilot. VASPilot consists of three main components: Web server (blue background box), CrewAI based multi agent cooperation core (gray background box) and Model Context Protocol type tool server (yellow backgound box).}
    \label{fig:overall}
\end{figure}

\section{\label{sec:methods}Methods}

\subsection{Overall structure of VASPilot}

The VASPilot program is primarily composed of three components: a web-based interface, a multi-agent collaborative core, and a tool server for the agents (Fig. \ref{fig:overall}). The multi-agent collaboration module is implemented using the CrewAI framework \cite{crewai_crewai_2024}, an open-source Python library for constructing cooperative AI agent teams. CrewAI facilitates the automation of complex tasks through modular workflows, role-specific agents, and flexible tool integration. The tool server is encapsulated within the Model Context Protocol (MCP)\cite{MCP2024}, a standardized interface for tool invocation and data provisioning. By adhering to MCP, our tools can be seamlessly incorporated into other frameworks with minimal additional effort.

Within this framework, we deploy one manager agent (orange box in Fig. \ref{fig:overall}) alongside three specialized worker agents (dashed box in Fig. \ref{fig:overall}): the Crystal Structure Agent, the VASP Agent, and the Result Validation Agent. In this work, all agents are powered by DeepSeek-V3-0324\cite{deepseek-ai_deepseek-v3_2025}, but they can be easily replaced with other LLMs in practice. The Crystal Structure Agent performs crystal‐structure searches, manipulations, and analyses. When the user did not specify a certain crystal structure file, the agent will search and download a structure from Materials Project \cite{Jain2013}. The VASP Agent orchestrates VASP calculations and produces the corresponding visualizations. The Result Validation Agent evaluates the accuracy and reliability of the computed outcomes.

Each worker agent maintains its own independent context and system prompt and can invoke only the subset of MCP-provided tools relevant to its tasks. In the case of the VASP Agent, we supplement the standard VASP-specific MCP tools with a custom CrewAI utility, 'wait\_calc\_tool'. This tool periodically polls the MCP server to determine whether ongoing calculations have finished, thereby mitigating potential network timeouts during long-running jobs.

When a user submits a mission via the web interface, the manager agent decomposes it into discrete tasks and dispatches each to the appropriate worker agent. Upon completion, each agent returns a detailed execution report to the manager. All outputs are archived in a Retrieval-Augmented Generation (RAG)\cite{lewis2021retrievalaugmentedgenerationknowledgeintensivenlp}-based memory pool: as an agent begins a task, it retrieves relevant historical entries to enrich its context and inform its decisions.

\subsection{Tools and Model Context Protocol Server}

\begin{figure}[h]
    \centering
\includegraphics[width=1\linewidth]{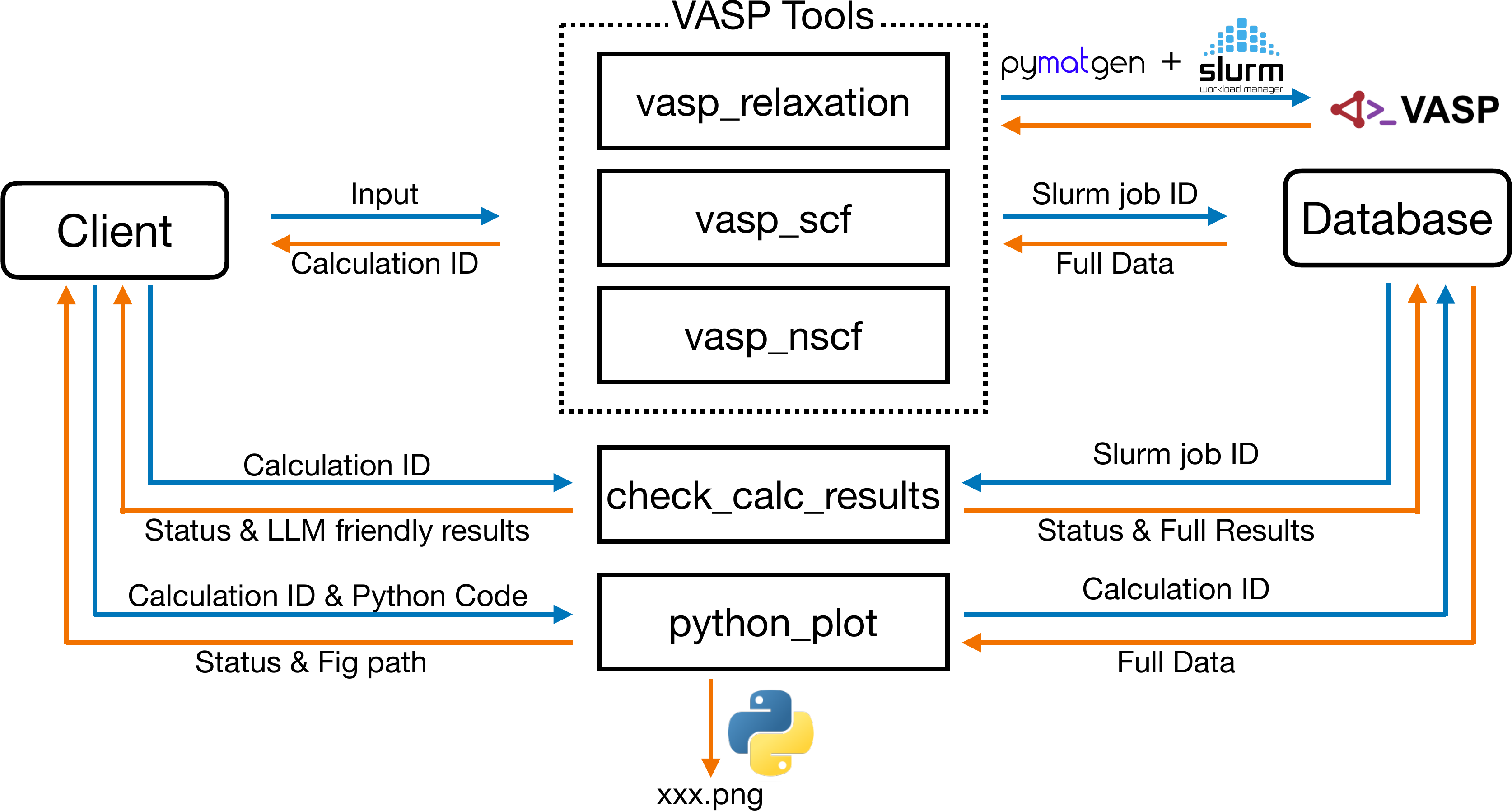}
    \caption{Specially tailored tools in VASPilot. VASP tools prepare input files and submits the calculation through pymatgen and slurm. 'check\_calc\_results' check the calculation status, read the results and store full data into database. 'python\_plot' reads full calculation data from database and execute the python code given by client. }
    \label{fig:Tools}
\end{figure}

In agentic LLM systems, tool design plays a pivotal role in ensuring both efficient performance and robust task execution. Because VASP generates extensive output that exceeds the native I/O capacity of most language models, we have implemented a coherent suite of utilities built on the pymatgen library to manage every stage of the VASP workflow, as shown in Fig. \ref{fig:Tools}.

First, our input-preparation tools (dashed box in Fig. \ref{fig:Tools}) automatically generate all necessary VASP input files and submit jobs to a Slurm scheduler. Each submission is tagged with a unique calculation ID, which, together with its Slurm job number, is recorded in a centralized database. To minimize communication overhead, only the calculation ID is returned to the agent. In the event of a workflow restart (for example, a non–self‐consistent band-structure run), the tool retrieves the relevant metadata and input files by calculation ID, reconstructs the working directory, and resubmits the job seamlessly.

Second, we provide a status-query utility that reports on both failed and successful calculations (check\_calc\_results tool in Fig. \ref{fig:Tools}). For failed jobs, the tool returns the VASP error messages; for successful runs, it reads the full result set via pymatgen, archives the data in the database, and extracts a concise summary of key properties (e.g., total energy, band gap, and Fermi energy) for downstream LLM consumption.

Finally, to enable flexible post-processing, we developed a plotting tool that executes client‐provided Python code(python\_plot tool in Fig. \ref{fig:Tools}). Given a list of calculation IDs and plotting instructions, the tool retrieves the corresponding datasets, runs the plotting code, and returns either the path to the generated image or any error messages encountered. This end-to-end framework offloads heavy I/O from the LLM, guarantees full traceability of every task, and supports reliable, error-tolerant computation and visualization.

\begin{figure*}
\centering
\includegraphics[width=\linewidth]{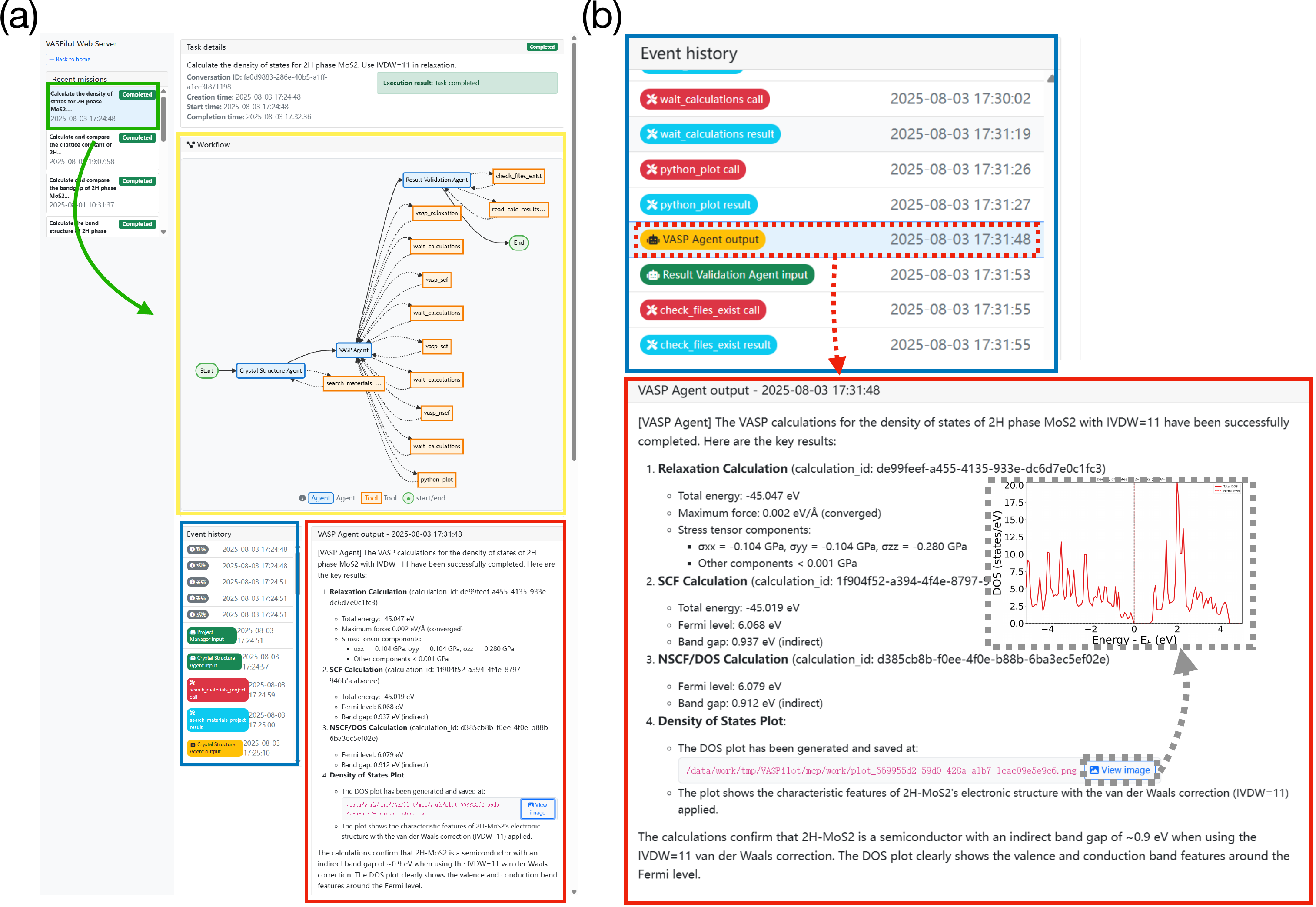}
    \caption{Web user interface of VASPilot. (a), The overview of the web user interface. By clicking the mission record in green box, the detail of mission will be shown on the right. The yellow box is the real-time workflow of the VASPilot. The blue box is the execution history of agents and tools. The red box is the details of selected execution history. (b), enlarged view of the execution history. The detailed information will be shown when user select an event. In detailed information area, figures and crystal structure files can be immediately shown by clicking the button. }
    \label{fig:UI}
\end{figure*} 

\subsection{Intuitive web interface}

Although VASPilot automates complex workflows, user oversight remains indispensable to ensure the accuracy of both intermediate steps and final outcomes. To facilitate this interaction, we have developed an intuitive, web-based interface using Python's Flask framework\cite{flask2025} (Fig. \ref{fig:UI}).

On the main page, users submit mission directives in natural language via a dedicated input dialog. Directly below this dialog, a chronological history of all previously submitted tasks is presented. By selecting any entry (green box in Fig. \ref{fig:UI} (a)), the user is navigated to a detailed task view in which a flowchart (yellow box in Fig. \ref{fig:UI} (a)) graphically represents the execution sequence and current status. Beneath the flowchart, expandable "Agent Execution" and "Tool Execution" logs (Fig. \ref{fig:UI} (b)) provide full visibility into each agent's and tool's inputs and outputs. Moreover, within these logs, crystal-structure visualizations and generated plots can be inspected in greater detail via clearly labeled buttons (Marked by gray dashed box in Fig. \ref{fig:UI}(b)). This design empowers users to both monitor overall progress at a glance and drill down into the computational data for rigorous verification.

\begin{figure*}
    \centering
    \includegraphics[width=1\linewidth]{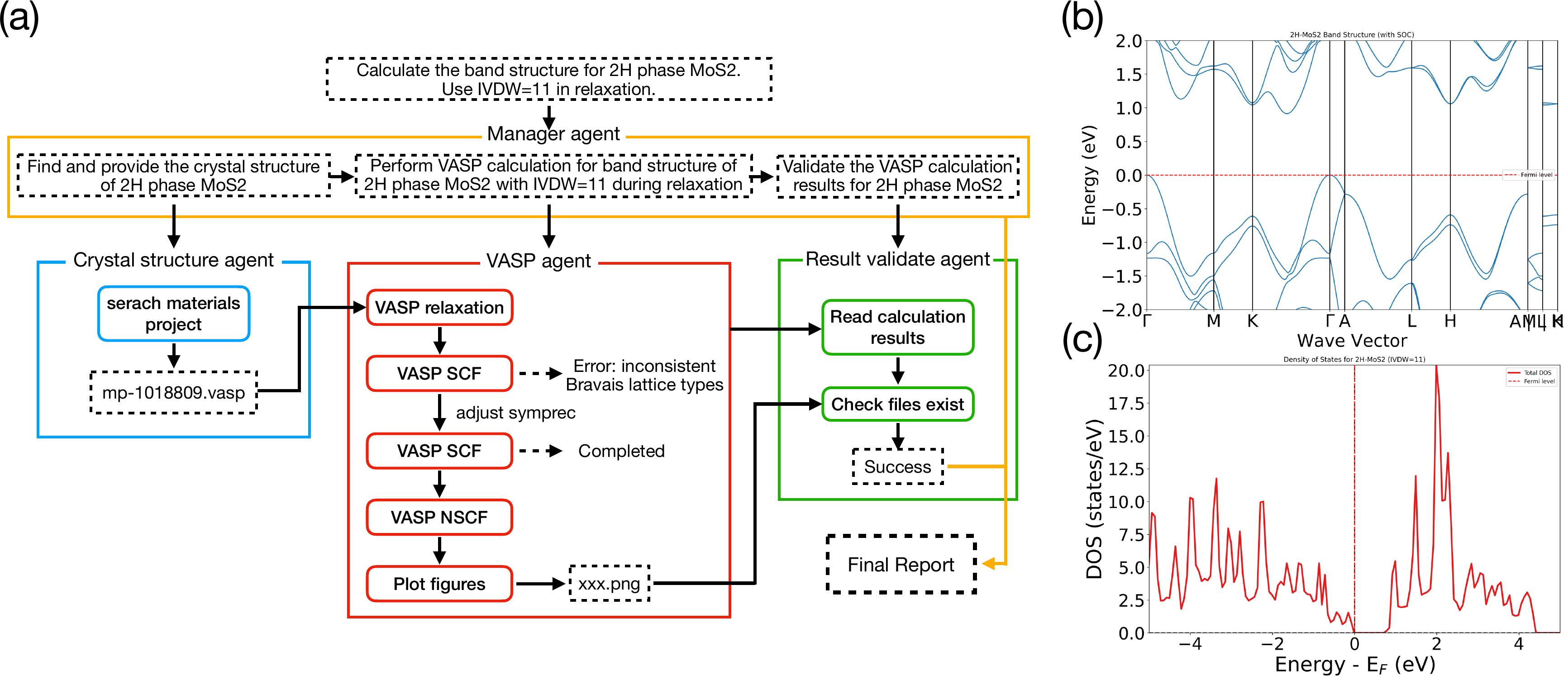}
    \caption{Band structure and density of states calculation of 2H-MoS$_2$ utilizing VASPilot. (a), workflow of band structure calculation. (b), Band structure figure calculated and plotted by VASPilot. (c), Density of states figure calculated and plotted by VASPilot. The workflow of Density of states calculation is almost the same as the band structure calculation. }
    \label{fig:BS_DOS_calc}
\end{figure*}

\begin{figure*}
    \centering
    \includegraphics[width=1\linewidth]{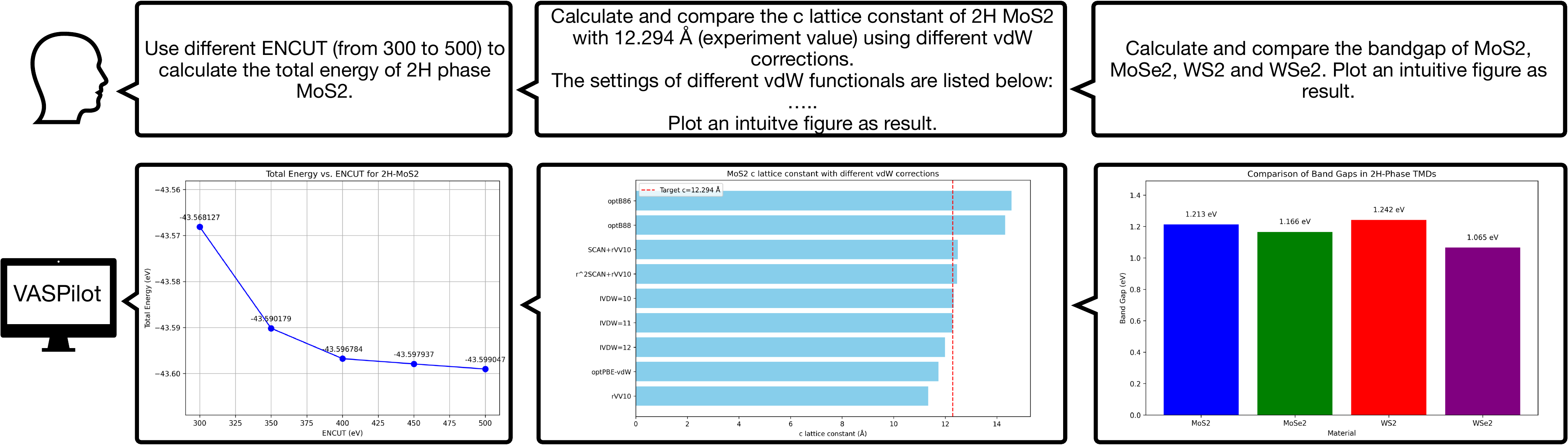}
    \caption{Prompts and results of three complex missions. The resulting figures are given by VASPilot in its final report. All these missions are finished without examples as context. In the second column, IVDW=10 is the DFT-D2 method of Grimme\cite{grimme_semiempirical_2006}. IVDW=11 is the DFT-D3 method of Grimme with zero-damping function\cite{grimme_consistent_2010}. IVDW=12 is the DFT-D3 method with Becke-Johnson damping function \cite{grimme_effect_2011}. IVDW=20 is the Tkatchenko-Scheffler method\cite{tkatchenko_accurate_2009}. optB86\cite{klimes_van_2011}, optB88\cite{klimes_chemical_2010}, SCAN+rVV10\cite{peng_versatile_2016}, r$^2$SCAN+rVV10\cite{ning_workhorse_2022}, rVV10\cite{sabatini_nonlocal_2013} and optPBE-vdW\cite{klimes_chemical_2010} are nonlocal vdW corrections. \label{fig:Harder}}
\end{figure*}

\section{\label{sec:results}Examples}

To evaluate VASPilot’s reliability, we performed band-structure and density-of-states (DOS) calculations for 2H-MoS$_2$, as illustrated in Fig. \ref{fig:BS_DOS_calc}. The workflow began when the Manager Agent instructed the Crystal Structure Agent to supply the crystal structure. This agent automatically retrieved the structure from the Materials Project database \cite{Jain2013} using the specified spacegroup. Following task delegation, the VASP Agent sequentially executed structural relaxation and self-consistent field (SCF) calculations by invoking requisite tools. During the initial SCF calculation, VASP returned a Bravais lattice inconsistency error. In response, VASPilot intelligently adjusted the symmetry tolerance parameter, resolving the issue and enabling successful completion of the subsequent SCF calculation. The VASP Agent then performed a non-self-consistent field (NSCF) band-structure calculation and generated the electronic band structure plot via built-in visualization tools. The Manager Agent subsequently directed the Result Validation Agent to verify the computational results. Acting autonomously, this agent confirmed the existence of the band structure image and the integrity of the computation using its dedicated validation tools. Finally, the Manager Agent synthesized all the results and compiled a comprehensive task report. The workflow of DOS calculation is almost the same as band structure calculation.

We further evaluated VASPilot's ability to execute novel, more complex workflows by defining three benchmark scenarios. First, we performed a plane-wave cutoff convergence test by calculating the total energy of 2H-MoS$_2$ over a range of ENCUT values from 300 eV to 500 eV. Second, we conducted structural relaxations of 2H-MoS$_2$ using several van der Waals correction schemes, then compared the optimized lattice constants against experimental measurements. Third, we computed and contrasted the band gaps of the 2H phases of MoS$_2$, MoSe$_2$, WS$_2$, and WSe$_2$. VASPilot successfully completed all three tasks. The detailed prompts and resulting data for these test cases are summarized in Fig. \ref{fig:Harder}.

These case studies underscore VASPilot's robustness and flexibility. The platform reliably executed standard workflows-such as electronic band-structure and density-of-states calculations-and, upon encountering VASP execution errors, automatically parsed the error messages, adjusted input parameters, and successfully resumed the runs. Moreover, even for complex missions lacking any contextual examples, VASPilot completed the tasks without intervention. Overall, these results demonstrate that VASPilot is a powerful and adaptable framework for fully automated DFT calculations.

\section{\label{sec:conclusions}Conclusions} 

In this work, we have developed VASPilot, an open‐source framework for fully automated VASP calculations. Built atop the CrewAI architecture and the Model Context Protocol (MCP), VASPilot seamlessly invokes specialized tools for structure retrieval and manipulation, input‐file generation, Slurm job orchestration, and postprocessing of results. We demonstrated its capability by executing automated band-structure and density-of-states workflows in which VASPilot parsed error messages, adapted input parameters on the fly, and successfully resumed interrupted runs. Beyond standard tasks, VASPilot handled more demanding benchmarks-converging total energies over a range of ENCUT values, optimizing lattice constants under different van der Waals correction schemes, and comparing band gaps across multiple 2H phase transition‐metal dichalcogenides.

These case studies attest to VASPilot's robustness and adaptability, enabling truly autonomous DFT workflows. Moreover, these capabilities can be readily extended through the integration of additional agents and tools, and the underlying framework can be easily adapted to other DFT codes by deploying the corresponding MCP server. We anticipate that VASPilot will streamline VASP-based research pipelines, reducing both the technical overhead and turnaround time for complex simulations. More broadly, we hope this work will inspire the development of similar agent-tool ecosystems for other computational platforms, accelerating automation across materials modeling and computational physics. 

\section{Data and Code availability}

The VASPilot source code and comprehensive documentation are publicly available under an open-source license on GitHub (\href{https://github.com/JiaxuanLiu-Arsko/VASPilot}{https://github.com/JiaxuanLiu-Arsko/VASPilot}), and an interactive demo can be accessed via the MaterialsGalaxy service portal (\href{https://cmpdc.iphy.ac.cn/materialsgalaxy/tools/vaspilot}{https://cmpdc.iphy.ac.cn/materialsgalaxy/tools/vaspilot}).

\begin{acknowledgments}
This work was supported by the Science Center of the National Natural Science Foundation of China (Grant No. 12188101), the National Key R\&D Program of China (Grant No. 2023YFA1607400, 2022YFA1403800), the National Natural Science Foundation of China (Grant No.12274436, 11925408, 11921004),  and  H.W. acoknowledge support from the New Cornerstone Science Foundation through the XPLORER PRIZE. 
\end{acknowledgments}

\bibliography{ref}

\end{document}